\documentstyle[12pt]{article}
\begin{document}

\def\emline#1#2#3#4#5#6{%
       \put(#1,#2){\special{em:moveto}}%
       \put(#4,#5){\special{em:lineto}}}

\begin{flushright}
Preprint SSU-HEP-97/03\\
Samara State University
\end{flushright}

\vspace{30mm}

\begin{center}
{\bf CONTRIBUTION OF HADRONIC VACUUM\\ POLARIZATION TO HYPERFINE\\
SPLITTING OF MUONIC HYDROGEN}\\
{\bf Faustov R.N.}\\
Scientific Council "Cybernetics" RAS,\\
117333, Vavilova, 40, Moscow, Russia,\\
{\bf Martynenko A.P.}\\
Samara State University, 443011, Ac. Pavlov, 1, Samara, Russia
\end{center}
\begin{abstract}
The contribution of hadronic vacuum polarization to the hyperfine splitting
of the muonic hydrogen ground state is calculated with the account of
experimental data on the cross section of $e^+e^-\rightarrow$ annihilation
into hadrons and the dipole parameterization of electromagnetic proton form
factors.

\end{abstract}

\newpage
Over many years the comparison of experimental and theoretical results
for hyperfine splitting of hydrogen, muonium and positronium ground
state energy level is considered as an important verification of our
understanding of the bound state problem in quantum electrodynamics.
Especially high precision was achieved at the present time when measuring
hydrogen atom hyperfine structure \cite{BY,DTF}:

\begin{equation}
\Delta E^{hf.s}_{exp.}=1420405.7517667 (9) KHz.
\end{equation}
Theoretical value of hyperfine splitting, which ordinary written in the form
\begin{equation}
\Delta E^{hf.s}_{th.}=\Delta E^F(1+\delta^{QED}+\delta^S+\delta^P),
~~~\Delta E^F=\frac{8}{3}\alpha^4\frac{\mu_P m^2_Pm^2_e}{(m_P+m_e)^3},
\end{equation}
($\mu_P$ is the proton magnetic moment, expressed in nuclear magnetons;
$m_P$, $m_e$ are proton and electron masses correspondingly),
contains besides electromagnetic correction $\delta^{QED}$ some more two
terms $\delta^S $ and $\delta^P$. These corrections take into account the influence
of strong interaction. The term $\delta^S$ describes the effects of proton
finite size and recoil contributions, and $\delta^P$ is the correction of
proton polarizability. Particle structure members $\delta^S$ and $\delta^P$ are
absent for purely leptonic systems \cite{Bodwin}. Electrodynamical
correction $\delta^{QED}$ for hyperfine structure of hydrogen atom, accounting
relativistic and radiative effects, is known at present time with the accuracy
0,12 ppm \cite{SGK}. The increase of accuracy in theoretical expression (2) is
impossible without the detailed consideration of strong interaction in the terms
$\delta^S$ and $\delta^P$. The calculation of different contributions to the
hydrogen atom hyperfine splitting (HFS), connected with proton structure
(correction $\delta^s)$) was done in the paper \cite{SGK}. Authors of \cite{Bakalov}
have proposed an experimental method to obtain more exact information about
proton electromagnetic structure. It based on the measurement of hyperfine
splitting of the muonic hydrogen ($\mu p$) ground state with the accuracy
$10^{-4}$. It is the authors opinion that such measurement will allow to find
the relation between the corrections $\delta^S$ and $\delta^P$, what will
increase the accuracy of theoretical formula (2). In the case of muonic
hydrogen, when the light electron must be replaced by the heavy muon, the
relative contribution of different effects in (2) is changed. So, the
hadronic vacuum polarization contribution, which was omitted in (2) in view of
it's small value by comparison with $\delta^S$ and $\delta^P$, may have more
relative contribution for muonic hydrogen HFS. The effect of hadronic vacuum
polarization must be considered in order to extract correctly the parameters
$\delta^S$ and $\delta^P$ from experiment. In this paper we have calculated
the hadronic vacuum polarization contribution to the hyperfine structure of
muonic hydrogen on the basis of quasipotential method \cite{Faustov}.
In this approach the particle interaction operator for $1\gamma$- and $2\gamma$-
processes takes the form \cite{MF}:
\begin{equation}
V=V_{1\gamma}+V_{2\gamma}=V^c+\Delta V,
\end{equation}
\begin{displaymath}
V_{1\gamma}=T_{1\gamma},~~~V_{2\gamma}=T_{2\gamma}-T_{1\gamma}\times G^f\times T_{1\gamma},  
\end{displaymath}
where $V^c$ is Coulomb potential, $T_{1\gamma}$, $T_{2\gamma}$ are one- and two-photon
scattering amplitudes beyond energy surface,
$[G^f]^{-1}=(b^2-\vec p^2)/2\mu_R$ is the free two-particle Green function
($\mu_R$ is relativistic reduced mass, $b^2$ is the square of particle relative
motion momentum on energy surface). Then we can write the displacement of energy
level as follows:
\begin{equation}
\Delta E=<\psi^c|\Delta V|\psi^c>,
\end{equation}
where $\psi^c$ is the ordinary Coulomb wave function. The hadronic vacuum
polarization contribution to the energy spectrum is determined by Feynman
diagrams represented on Fig.1 by means (3). To take into account vacuum polarization
in these one-loop diagrams, we must do the following substitution in photon
propagator \cite{t4}
\begin{equation}
\frac{1}{k^2+i\varepsilon}\rightarrow\left(\frac{\alpha}{\pi}\right)^2\int_
{s_{th.}}^\infty\frac{\rho(s)ds}{k^2-s+i\varepsilon},
\end{equation}
where the spectral function $\rho(s)$ is connected with known cross section of
$e^+e^-$ - annihilation to hadrons $\sigma^h$:
\begin{equation}
\rho(s)=\frac{R(s)}{3s}=\frac{\sigma^h(e^+e^-\rightarrow hadrons)}{3s\sigma_{\mu\mu}
(e^+e^-\rightarrow\mu^+\mu^-)},
\end{equation}
and $\sigma_{\mu\mu}(e^+e^-\rightarrow\mu^+\mu^-)=4\pi\alpha^2/3s$ is the
$e^+e^-$- annihilation cross section to muonic pair.

\begin{figure}
\unitlength=1.00mm
\special{em:linewidth 1pt}
\linethickness{1pt}
\begin{picture}(119.00,80.00)
\emline{17.00}{80.00}{1}{20.00}{77.00}{2}
\emline{20.00}{77.00}{3}{14.00}{74.00}{4}
\emline{14.00}{74.00}{5}{20.00}{71.00}{6}
\emline{20.00}{71.00}{7}{14.00}{68.00}{8}
\emline{14.00}{68.00}{9}{20.00}{65.00}{10}
\emline{20.00}{65.00}{11}{14.00}{62.00}{12}
\emline{14.00}{62.00}{13}{20.00}{59.00}{14}
\emline{20.00}{59.00}{15}{14.00}{56.00}{16}
\emline{14.00}{56.00}{17}{20.00}{53.00}{18}
\emline{20.00}{53.00}{19}{14.00}{50.00}{20}
\emline{14.00}{50.00}{21}{20.00}{47.00}{22}
\emline{20.00}{47.00}{23}{14.00}{44.00}{24}
\emline{14.00}{44.00}{25}{20.00}{41.00}{26}
\emline{20.00}{41.00}{27}{14.00}{38.00}{28}
\emline{14.00}{38.00}{29}{20.00}{35.00}{30}
\emline{20.00}{35.00}{31}{14.00}{32.00}{32}
\emline{14.00}{32.00}{33}{20.00}{29.00}{34}
\emline{20.00}{29.00}{35}{14.00}{26.00}{36}
\emline{14.00}{26.00}{37}{20.00}{23.00}{38}
\emline{20.00}{23.00}{39}{17.00}{20.00}{40}
\emline{47.00}{80.00}{41}{50.00}{77.00}{42}
\emline{50.00}{77.00}{43}{44.00}{74.00}{44}
\emline{44.00}{74.00}{45}{50.00}{71.00}{46}
\put(47.00,50.00){\circle{14.00}}
\emline{50.00}{71.00}{47}{44.00}{68.00}{48}
\emline{44.00}{68.00}{49}{50.00}{65.00}{50}
\emline{50.00}{65.00}{51}{44.00}{62.00}{52}
\emline{44.00}{62.00}{53}{50.00}{59.00}{54}
\emline{50.00}{59.00}{55}{47.00}{57.00}{56}
\emline{47.00}{20.00}{57}{50.00}{23.00}{58}
\emline{50.00}{23.00}{59}{44.00}{26.00}{60}
\emline{44.00}{26.00}{61}{50.00}{29.00}{62}
\emline{50.00}{29.00}{63}{44.00}{32.00}{64}
\emline{44.00}{32.00}{65}{50.00}{35.00}{66}
\emline{50.00}{35.00}{67}{44.00}{38.00}{68}
\emline{44.00}{38.00}{69}{50.00}{41.00}{70}
\emline{50.00}{41.00}{71}{47.00}{43.00}{72}
\emline{105.00}{80.00}{73}{108.00}{77.00}{74}
\emline{108.00}{77.00}{75}{102.00}{74.00}{76}
\emline{102.00}{74.00}{77}{108.00}{71.00}{78}
\emline{108.00}{71.00}{79}{102.00}{68.00}{80}
\emline{102.00}{68.00}{81}{108.00}{65.00}{82}
\emline{108.00}{65.00}{83}{102.00}{62.00}{84}
\emline{102.00}{62.00}{85}{108.00}{59.00}{86}
\emline{108.00}{59.00}{87}{102.00}{56.00}{88}
\emline{102.00}{56.00}{89}{108.00}{53.00}{90}
\emline{108.00}{53.00}{91}{102.00}{50.00}{92}
\emline{102.00}{50.00}{93}{108.00}{47.00}{94}
\emline{108.00}{47.00}{95}{102.00}{44.00}{96}
\emline{102.00}{44.00}{97}{108.00}{41.00}{98}
\emline{108.00}{41.00}{99}{102.00}{38.00}{100}
\emline{102.00}{38.00}{101}{108.00}{35.00}{102}
\emline{108.00}{35.00}{103}{102.00}{32.00}{104}
\emline{102.00}{32.00}{105}{108.00}{29.00}{106}
\emline{108.00}{29.00}{107}{102.00}{26.00}{108}
\emline{102.00}{26.00}{109}{108.00}{23.00}{110}
\emline{108.00}{23.00}{111}{105.00}{20.00}{112}
\emline{75.00}{80.00}{113}{78.00}{77.00}{114}
\emline{78.00}{77.00}{115}{72.00}{74.00}{116}
\emline{72.00}{74.00}{117}{78.00}{71.00}{118}
\put(75.00,50.00){\circle{14.00}}
\emline{78.00}{71.00}{119}{72.00}{68.00}{120}
\emline{72.00}{68.00}{121}{78.00}{65.00}{122}
\emline{78.00}{65.00}{123}{72.00}{62.00}{124}
\emline{72.00}{62.00}{125}{78.00}{59.00}{126}
\emline{78.00}{59.00}{127}{75.00}{57.00}{128}
\emline{75.00}{20.00}{129}{78.00}{23.00}{130}
\emline{78.00}{23.00}{131}{72.00}{26.00}{132}
\emline{72.00}{26.00}{133}{78.00}{29.00}{134}
\emline{78.00}{29.00}{135}{72.00}{32.00}{136}
\emline{72.00}{32.00}{137}{78.00}{35.00}{138}
\emline{78.00}{35.00}{139}{72.00}{38.00}{140}
\emline{72.00}{38.00}{141}{78.00}{41.00}{142}
\emline{78.00}{41.00}{143}{75.00}{43.00}{144}
\emline{10.00}{80.00}{145}{55.00}{80.00}{146}
\emline{55.00}{20.00}{147}{10.00}{20.00}{148}
\emline{68.00}{80.00}{149}{112.00}{80.00}{150}
\emline{112.00}{20.00}{151}{68.00}{20.00}{152}
\put(62.00,50.00){\makebox(0,0)[cc]{+}}
\put(112.00,51.00){\makebox(0,0)[cc]{+}}
\put(5.00,80.00){\makebox(0,0)[cc]{p}}
\put(5.00,20.00){\makebox(0,0)[cc]{$\mu$}}
\put(118.00,80.00){\makebox(0,0)[cc]{p}}
\put(118.00,20.00){\makebox(0,0)[cc]{$\mu$}}
\put(119.00,71.00){\makebox(0,0)[lc]{diagrams}}
\put(119.00,61.00){\makebox(0,0)[lc]{with crossed}}
\put(119.00,51.00){\makebox(0,0)[lc]{photon lines}}
\put(119.00,41.00){\makebox(0,0)[lc]{and subtrahend}}
\put(119.00,31.00){\makebox(0,0)[lc]{diagrams}}
\put(17.00,80.00){\circle*{5.20}}
\put(47.00,80.00){\circle*{5.20}}
\put(75.00,80.00){\circle*{5.20}}
\put(105.00,80.00){\circle*{5.20}}
\end{picture}
\caption{Feynman diagrams, defining the hadronic vacuum polarization
contribution to hyperfine structure of atom ($\mu p$)}
\end{figure}
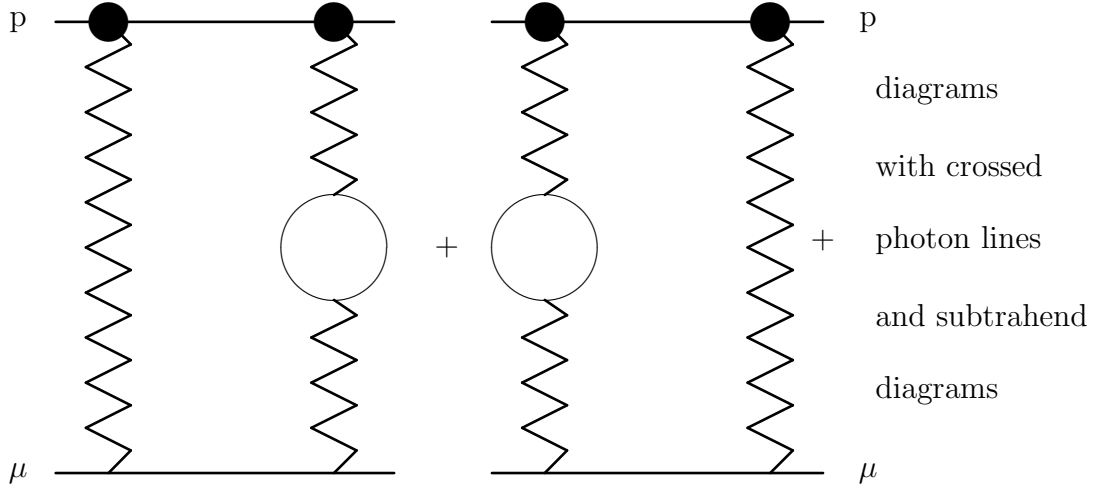

Let consider the proton factor, corresponding to two-photon exchange diagrams:
\begin{equation}
M_p^{\mu\nu}=\frac{\Gamma^\mu(\gamma^0E_2-\hat p+m_p)\Gamma^\nu}
{p^2+b^2-2E_2p_0+i\varepsilon}+
\frac{\Gamma^\nu(\gamma^0E_2+\hat p+m_p)\Gamma^\mu}
{(p+t)^2+b^2+2E_2p_0+i\varepsilon},
\end{equation}
where $p^\mu$ is the loop 4-momentum, 4-momentum $t=(0,\vec p-\vec q)$ is
determined by relative motion momenta $\vec p$, $\vec q$ of initial and final
states correspondingly. Proton vertex function $\Gamma^\mu$ may be expressed
through Dirac $\rho(p^2)$ and Pauli $f(p^2)$ form factors, describing the
proton electromagnetic structure:
\begin{equation}
\Gamma^\mu=\left[\gamma^\mu\rho(p^2)+\frac{i}{2m_p}\sigma^{\mu\nu}p_\nu f(p^2)
\right],
\end{equation}
where $\rho(0)=e,~~f(0)=e\kappa$, $\kappa$ is the proton anomalous magnetic moment:
$\kappa$=1.792847386(63) \cite{BY}. Muonic tensor of considered one-loop amplitudes
is equal:
\begin{equation}
M^{\mu\nu}_{muon}=\frac{\gamma^\mu(\gamma^0E_1+\hat p+m_p)\gamma^\nu}
{p^2+b^2+2E_1p_0+i\varepsilon}.
\end{equation}
The hadronic vacuum polarization contribution to the ground state hyperfine
splitting of muonium and positronium was obtained in \cite{STY}. It is clear
that for exact calculation of discussed one-loop corrections in ($\mu p$),
we must consider the dependence of form factors $\rho(p^2)$, $f(p^2)$
from loop momentum p. To do numerical calculations of the corrections
$O(\alpha^6)$ caused by hadronic vacuum polarization we can use dipole
parameterization of proton electromagnetic form factors of the form:
\begin{equation}
\rho(p^2)=\frac{G_E-\frac{p^2}{4m^2_p}G_M}{1-\frac{p^2}{4m^2_p}},~~~
f(p^2)=\frac{G_M-G_E}{1-\frac{p^2}{4m^2_p}},
\end{equation}
\begin{equation}
G_M=\frac{1+\kappa}{\left(1-\frac{p^2}{\Lambda^2}\right)^2},~~~
G_E=\frac{1}{\left(1-\frac{p^2}{\Lambda^2}\right)^2},
\end{equation}
where the parameter of proton structure was taken in the kind $\Lambda=0,898 m_p$
\cite{BY}. Multiplying the expressions (7) and (9) for construction of
quasipotential, we obtain the following sum of form factor terms with the
accuracy $O(\alpha^6)$:
\begin{displaymath}
<M^{\mu\nu}_pM^{\mu\nu}_{muon}>=-\rho^2(p^2)<\gamma^\mu\hat p\gamma^\nu>
<\gamma^\mu\hat p\gamma^\nu>\left(\frac{1}{D_p(-p)}+\frac{1}{D_p(p)}\right)
\frac{1}{D_\mu(p)}-
\end{displaymath}
\begin{displaymath}
-2p^2\frac{\rho(p^2)f(p^2)}{m_p}<\gamma^\mu\gamma^\nu>
<\gamma^\mu\hat p\gamma^\nu>\left(\frac{1}{D_p(-p)}+\frac{1}{D_p(p)}\right)
\frac{1}{D_\mu(p)}+
\end{displaymath}
\begin{displaymath}
+4\rho(p^2)f(p^2)<\hat p\gamma^\mu>
<\hat p\gamma^\mu>\left(\frac{1}{D_p(-p)}+\frac{1}{D_p(p)}\right)
\frac{1}{D_\mu(p)}-
\end{displaymath}
\begin{equation}
-\frac{f^2(p^2)}{m_p^2}p^2<\gamma^\mu\hat p\gamma^\nu>
<\gamma^\mu\hat p\gamma^\nu>\left(\frac{1}{D_p(-p)}+\frac{1}{D_p(p)}\right)
\frac{1}{D_\mu(p)}+
\end{equation}
\begin{displaymath}
+\left(\frac{f(p^2)}{m_p}\right)^2 2m_pp^2<\gamma^\mu\gamma^\nu>
<\gamma^\mu\hat p\gamma^\nu>\left(\frac{1}{D_p(-p)}-\frac{1}{D_p(p)}\right)
\frac{1}{D_\mu(p)}+
\end{displaymath}
\begin{displaymath}
+\left(\frac{f(p^2)}{m_p}\right)^2 2m_pp^0<\gamma^\mu\hat p\gamma^\nu>
<\gamma^\mu\hat p\gamma^\nu>\left(\frac{1}{D_p(-p)}-\frac{1}{D_p(p)}\right)
\frac{1}{D_\mu(p)},  
\end{displaymath}
where
\begin{displaymath}
D_p(\pm p)=p^2\pm2m_pp_0+i\varepsilon,~~~D_\mu(p)=p^2+2m_\mu p_0+i\varepsilon.
\end{displaymath}
To make numerical calculations let transform (12) to convenient form. First
of all we pick out in proton factor $1/D_p(-p)$ of crossed two-photon diagram
the separate term with $\delta(p_0)$ function, which cancelled in future
calculations by similar addendum of iteration part of quasipotential (3):
\begin{equation}
\frac{1}{D_p(-p)}=\frac{1}{2(m_p+m_\mu)}\left[-2\pi i\delta(p_0)-\frac{1}{p_0+i\varepsilon}-
\frac{D_\mu(p)}{(-p_0+i\varepsilon)D_p(-p)}\right].
\end{equation}
It is useful also to represent the other proton factor $1/D_p(p)$, appearing in (12),
in the following manner:
\begin{equation}
\frac{1}{D_p(p)}=\frac{1}{2(m_p-m_\mu)(p_0+i\varepsilon)}\left[1-\frac{D_\mu(p)}
{D_p(p)}\right].
\end{equation}
Accounting these relations, we may transform different parts of (12) as
follows:
\begin{equation}
\frac{p_0}{D_\mu(p)}\left[\frac{1}{D_p(-p)}-\frac{1}{D_p(p)}\right]=
\frac{m_p}{m_p^2-m_\mu^2}\left[\frac{1}{D_p(p)}-\frac{1}{D_\mu(p)}\right],
\end{equation}

\begin{equation}
\frac{p_0}{D_\mu(p)}\left[\frac{1}{D_p(-p)}+\frac{1}{D_p(p)}\right]=
\frac{m_\mu}{m_p^2-m_\mu^2}\left[\frac{1}{D_\mu(p)}-\frac{1}{D_p(p)}\right],
\end{equation}
  
\begin{equation}
\frac{p^2}{D_\mu(p)}\left[\frac{1}{D_p(-p)}+\frac{1}{D_p(p)}\right]\approx
\frac{2}{m_p^2-m_\mu^2}\left[\frac{m_p^2}{D_p(p)}-\frac{m_\mu^2}{D_\mu(p)}\right],
\end{equation}
After that we rotate the $p_0$ contour to the imaginary axis and call the new
variable $\xi$. The integrated factors are modified as follows:
\begin{equation}
\frac{1}{p^2+i\varepsilon}\rightarrow -\frac{1}{\vec p^2+\xi^2},~~~
\frac{1}{D_p(p)}\rightarrow \frac{-\vec p^2+\xi^2-2im_p\xi}{(\vec p^2+\xi^2)^2+4m_p^2\xi^2}.
\end{equation}
As a result of such transformations we obtain that the contribution of 1, 3 and 4
form factor terms in (12) to muonic hydrogen HFS with the account of
hadronic vacuum polarization is determined by triple integral of the following
form:
\begin{equation}
\Delta E^{hf.s}_1=\tilde E_F\frac{m_p m_\mu}{m_p^2-m_\mu^2}\frac{32\alpha^2}{\pi^2}
\int_{s_0}^\infty \rho(s)ds\int_0^\infty d\xi\int_0^\infty p^2 dp
\end{equation}
\begin{displaymath}
\frac{[\rho^2-\frac{f^2}{m_p^2}(\vec p^2+\xi^2)](3\xi^2+2\vec p^2)+4\vec p^2\rho f}
{(\vec p^2+\xi^2)(\vec p^2+\xi^2+s)}\left[\frac{m_p^2}{(\vec p^2+\xi^2)^2+4m_p^2\xi^2}-
\frac{m_\mu^2}{(\vec p^2+\xi^2)^2+4m_\mu^2\xi^2}\right].
\end{displaymath}
To do integration over s, we have used the parameterization of
$\rho(s)$ by means of new experimental data on annihilation cross section
$e^+e^-\rightarrow hadrons$ \cite{Dubn,Swartz,Eid,Isaev}. The main
contribution to the cross section $\sigma^h$ is determined by the process
$e^++e^-\rightarrow \pi^++\pi^-$. The cross section of this reaction is
proportional to the squared module of $\pi$ meson form factor $F_\pi$.
Supposing that the $\rho$ exchange diagram gives the basic contribution
to $F_\pi$, we have used the following well-known expression of pionic
form factor, taking into account the contribution of $\rho-\omega$
interference term \cite{Dubn,Isaev}:
\begin{equation}
F_\pi(s)=\frac{m^2_\rho(1+d\Gamma_\rho/m_\rho)}{(m_\rho^2-s+i m_\rho\Gamma_\rho
(p/p_0)^3m_\rho/\sqrt{s}}+\zeta e^{i\phi}\frac{m_\omega^2}{m_\omega^2-s+i m_\omega
\Gamma_\omega},
\end{equation}
where the $\rho$ meson decay width $\Gamma_\rho$=0.15 Gev, $\rho$ meson mass
$m_\rho$=0.77 Gev and d=0.5. The effects of $\rho-\omega $ interference
are described by the parameters from \cite {Dubn}. Substituting (20) to the
spectral function
\begin{equation}
\rho_{\pi\pi}(s)=\frac{(s-4m_\pi^2)^{3/2}}{12s^{5/2}}|F_\pi(s)|^2,
\end{equation}
we obtain from (19) the corresponding contribution to HFS of muonic
hydrogen:
\begin{equation}
\Delta E^{hf.s}_{h~1}(\rho\rightarrow 2\pi)=402.40~~MHz,~~~~\tilde E^F=15796.2~~GHz.
\end{equation}
The term of (19), proportional to $f^2/m_p^2$ leads to the negative
contribution 23,84 MHz in the energy spectrum. It is interesting to compare
the quantity (19) with the result of it calculation in the point-like proton
approximation, when we assume $\rho (k^2)\approx\rho (0)=e$,
$f(k^2)\approx f(0)=e\kappa$. Keeping in mind that the main contribution in
(19) is determined by first and third terms of (12), let transform their
quantity in the hyperfine splitting of ($\mu p$), fulfilling analytical integration
over loop momentum p by means of Feynman parameterization. As a result we find:

\begin{equation}
\Delta E^{hf.s}_h=\left(\frac{\alpha}{\pi}\right)^2\frac{m_\mu m_P}{m^2_P-m^2_\mu}
\tilde E_F\int_{4m^2_\pi}^\infty ds\rho(s)\int_0^1 dx\left[S(m_P,x)-
S(m_\mu,x)\right],
\end{equation}
where
\begin{displaymath}
S(m,x)=\frac{16-6x-x^2+4\kappa(8-6x+x^2)}{x^2+\frac{s(1-x)}{m^2}},
~~~\tilde E^F=8\alpha^4m^2_\mu m^2_P/(m_P+m_\mu)^3.
\end{displaymath}

Integration over parameter x may be done analytically also, and to integrate
over s, we have used parameterization of $\rho(s)$ in the form (21). Then the
contribution of first and third terms of (12) to HFS in the point-like
proton approximation is equal to:
\begin{equation}
\Delta E_h^{hf.s}(\rho\rightarrow 2\pi)=882.95~~MHz.
\end{equation}
Comparing this value with the result (22), we may state that using of form
factors (10), (11) in vertex operator (8) leads to essential expected
decrease of corresponding correction by comparison to approximation
$\rho(p^2)\approx\rho(0)$, $f(p^2)\approx f(0)$. The other parts of (12)
give the contribution to HFS, which may be represented as a sum of two
integrals similar to (19):
\begin{equation}
\Delta E^{hf.s}_2=\tilde E_F\frac{m_p m_\mu}{m_p^2-m_\mu^2}\frac{32\alpha^2}{\pi^2}
\int_{s_0}^\infty \rho(s)ds\int_0^\infty d\xi\int_0^\infty p^4 dp
\end{equation}
\begin{displaymath}
\frac{2f^2(p^2)}{(\vec p^2+\xi^2+s)}
\left[\frac{1}{(\vec p^2+\xi^2)^2+4m_\mu^2\xi^2}-
\frac{1}{(\vec p^2+\xi^2)^2+4m_p^2\xi^2}\right],
\end{displaymath}

\begin{equation}
\Delta E^{hf.s}_3=\tilde E_F\frac{m_p m_\mu}{m_p^2-m_\mu^2}\frac{192\alpha^2}{\pi^2}
\int_{s_0}^\infty \rho(s)ds\int_0^\infty d\xi\int_0^\infty p^2 dp
\end{equation}
\begin{displaymath}
\frac{\rho(p^2)f(p^2)m_\mu(\vec p^2+\xi^2)}
{m_p(\vec p^2+\xi^2+s)}\left[\frac{1}{(\vec p^2+\xi^2)^2+4m_\mu^2\xi^2}-
\frac{1}{(\vec p^2+\xi^2)^2+4m_p^2\xi^2}\right].
\end{displaymath}

We have represented in the table the total contribution of all form factor
terms of (12) with function $\rho_{\pi\pi}(s)$ to the HFS of muonic hydrogen,
which is formed by the expressions (19), (25) and (26). Let  take into account
the contributions of $\omega$ and $\phi$ mesons to hyperfine structure of ($\mu p)$.
To do this would require Breit-Wigner representation \cite{Isaev} of spectral
function $\rho(s)$:
\begin{equation}
\rho_{\omega,\phi}=\frac{m_{\omega,\phi}\Gamma_{\omega,\phi}}{4sf_{\omega,\phi}
[4(\sqrt{s}-m_{\omega,\phi})^2+\Gamma^2_{\omega,\phi}]},~~~f_{\omega,\phi}=
\frac{\alpha^2m_{\omega,\phi}}{12\Gamma^{ee}},
\end{equation}
where $m_\omega=0.782$ Gev, $m_\phi=1.019$ Gev, $\Gamma_\omega=8.43$ Mev,
$\Gamma_{\phi}=4.43$ Mev, $\Gamma^{ee}_{\omega}=0.60$ Kev, $\Gamma^{ee}_\phi=
1.37$ Kev \cite{PDG}. Then after calculation of triple integrals (19) we
obtain the summary contribution of the following kind:
\begin{equation}
\Delta E^{hf.s}_h(\omega)=45.74~~MHz,
\end{equation}
\begin{equation}
\Delta E^{hf.s}_h(\phi)=48.97~~MHz.
\end{equation}

Spectral density R(s) was parameterized in the energy range $0.64\leq s\leq 1.44$,
between $\omega$ and $ \phi$ resonances on the basis of present experimental
data \cite{Eid} on $e^+e^-$ annihilation cross section in the form:
$R(s)=A+Bs+Cs^2$ (A=22.68, B=-37.72, C=15.84).
To obtain the contributions of $J/\Psi$ and $\Upsilon$ particle families
we have also used the Breit-Wigner formula (27) with known experimental values
for total and partial decay widths of vector resonances to $e^+e^-$ pair
\cite{PDG}. The calculation of background contribution at $s\geq 1,44$ was
performed as in the paper \cite{KF}. The full integration region over s was
divided on six intervals. In each interval the spectral density was founded
by fitting of experimental points after subtraction of resonance contribution.
Our numerical calculations show, that the main part of hadronic vacuum
polarization contribution to hyperfine splitting of muonic hydrogen caused
by $\pi$ meson form factor (20). Theoretical error, not exceeding $10\%$,
reflects spread in the choice of fitting parameters for cross section
$\sigma(e^+e^-\rightarrow hadrons)$ and for nucleon form factors. So, we
actually see that whereas for hydrogen atom the relative contribution of
hadronic vacuum polarization to ground state hyperfine splitting is about
$10^{-8}$ \cite{SGK}, for muonium -  $5.6\cdot 10^{-8}$ \cite{STY,KF}, then
in the case of muonic hydrogen it increases to $19.7\cdot 10^{-6}$. The
essential modification of hadronic vacuum polarization relative contribution
to HFS of muonic hydrogen is connected with increase of particle relative
motion momentum, which achieves the values of order of hadronic masses.
Obtained in this paper value of correction $\Delta E^{hf.s}_h=861.05~~MHz$,
must be accounted for future comparison with experimental result and for
more exact determination of $\delta^S$ and $\delta^P$ \cite{Bakalov}.

When constructing two-photon interaction quasipotential we take into account
in the virtual Compton scattering amplitudes only elastic proton intermediate
states. We have investigated exactly this mechanism of the hadronic vacuum
polarization contribution to HFS of muonic hydrogen, which gives the
correction to $\delta^S$. It is possible additional "nonelastic" process
considering the excitement in the intermediate states of "ladder" diagrams of
$\Delta $ isobar or other nucleon resonances (correction $\delta^P$).
The contribution of such nonelastic mechanism demands special study.

\vspace{10mm}

\begin{tabular}{|c|c|}	\hline
Interval $\sqrt{s}$,~~Gev & $\Delta E^{hf.s}_h~~~(\mu p)$, MHz	  \\ \hline
$\rho$ & 442.97 \\
$\omega$ & 45.74  \\
$\phi  $ & 48.97  \\
$J/\Psi$ family  &  11.04  \\
$\Upsilon$ family  &  0.11  \\
Background    &  \\
$0.8\leq\sqrt{s}\leq 1.2$~~~$R=A+Bs+Cs^2$	 &  131.90  \\
$1.2\leq\sqrt{s}\leq 3.0$~~~$R=2.843s^{-0.181}$  &  104.01 \\
$3.0\leq\sqrt{s}\leq 7.4$~~~$R=1.293s^{0.692}$	&  70.78  \\
$7.4\leq\sqrt{s}\leq 12.0$~~~R=3.91  &	3.46  \\
$12.0\leq\sqrt{s}\leq 36.0$~~~R=4.01  &  1.94  \\
$36.0\leq\sqrt{s}\leq 47.0$~~~R=4.13  &  0.11  \\
$\sqrt{s}\geq 47.0$~~~QCD  &  0.02  \\	\hline
Summary contribution   &  861.05   \\	\hline
\end{tabular}

\vspace{10mm}

Nevertheless, it is possible to give the following approximate appreciation
of hadronic vacuum polarization contribution to HFS of muonic hydrogen,
taking into account proton polarizability. In the case of ordinary hydrogen
the polarization correction $\delta^P$ in (2) is no more than 4 ppm [4].
It is equal approximately to $10\%$ from elastic proton contribution
$\delta^S$. Then we can obtain the approximate value of discussed contribution
from proton polarizability replacing electron mass $m_e$ by muon mass
$m_\mu$ (as it follows from (23)) and adding one degree of $\alpha$ for
calculation of vacuum polarization. So, in the case of muonic hydrogen
such "nonelastic" mechanism may give contribution to HFS of order of
4 ppm$\frac{m_\mu}{m_e}\alpha\approx 6.0$ ppm. This value is about 30\%
from relative contribution of hadronic vacuum polarization, obtained in this
paper and equal to $19.7 ppm$.  

We are grateful to F.~Jegerlehner, A.~Karimkhodzaev,
S.~Karshenboim and V.~Saleev for useful discussions. This work was
supported by the Russian Foundation for Basic Research (grant no. 96-02-17309).

\end{document}